\documentstyle[prl,aps]{revtex}
\begin{document}
\draft

\title{Non-deterministic Chaos}

\author{D. D. Dixon}
\address{
Institute for Geophysics and Planetary Physics\\
University of California \\
Riverside, CA 92521\\
DIXON@UCRPHF.UCR.EDU
}

\date{\today}

\maketitle

\begin{abstract}
Non-deterministic chaos is a form of low-dimensional dynamics
which is characterized by
the existence of a countable set of {\em sensitive decision
points} (SDP's).  Away from these points, the dynamics is well-behaved.
Near these points, however, perturbations (e.g., thermal noise)
may cause the
outgoing trajectory to be chosen randomly.
An example of a non-determinstic chaotic system is given, and a
statistical method of analyzing the resultant dynamics is developed.
\end{abstract}


\paragraph*{Introduction}

Solutions of non-linear ordinary differential equations (ODE's) sometimes
exhibit the phenomenon of {\em deterministic chaos}, displaying sensitive
dependence on initial conditions and long-term
unpredictability~\cite{rasband}.  The
term ``deterministic'' is used because the ODE's are described by
functions of the dynamical variables and time which are well-defined
in the dynamical domain.  An oft-cited examples is the kicked harmonic
oscillator, which may be described by the equation
\begin{equation}\label{eq:kho}
\ddot{x} + \omega^2 x = A f(x,t) \sum_{n=1}^{\infty} \delta(t - nT),
\end{equation}
where $\omega$ is the frequency of the oscillator, $A$ is some constant,
and $f(x,t)$ is some arbitrary well-defined function.  The solution
of Eq.\ (\ref{eq:kho}) is simply that of a simple harmonic oscillator
(an ellipse in the phase plane) until the kick at $t = nT$ occurs.  At
this point the solution jumps from one ellipse to another as determined
by the magnitude $A f(x, nT)$ of the kick at this point.
For an appropriate
choice of $A$ and $f(x,t)$, the solution $x(t)$ of Eq.\ (\ref{eq:kho})
is chaotic~\cite{rasband}.\\

Let us now suppose that $f(x,t)$ is {\em not} well-defined, but rather is
a random function of time.  By ``random'', we mean that any point on
this function has zero correlation with all other points on the function.
The effect of this is to give the oscillator an unpredictable kick
whenever $t = nT$.  The solutions will thus be similar to those described
above, with the important difference that the kick-induced jump from one
phase plane ellipse to another is non-deterministic, and cannot be
predicted.\\

We will not examine the particular case of the randomly kicked harmonic
oscillator (RKHO) here.  We note, however, that the random function $f(x,t)$ is
expected to describable by some distribution about a mean value
$\langle f(x,t) \rangle$.  If this distribution is everywhere $> 0$, we
are then led to the following observations:
\begin{enumerate}
\item{The dynamical trajectory after the kick at $t = nT$ is completely
uncorrelated with the trajectory before.}
\item{We may make {\em statistical} predictions of what will occur after
each kick based on the distribution of $f(x,t)$.  From this, we should
be able to find some probability density that the dynamical trajectory
will visit a particular point in phase space.}
\end{enumerate}

\paragraph*{Realization}
The randomly kicked harmonic oscillator, while providing a useful
illustrative example, requires a random driving force of large
amplitude, which seems rather contrived.  We seek a more general
paradigm where seemingly insignificant random fluctuations can induce
the type of non-determinism seen above.  To this end we design a
``toy'' dynamical system, whose orbits are all circles parameterized
by their radius, $r$.  The key feature of our toy system is that
all of these circles share a common tangent point at the origin
(see Fig.~\ref{fig:circles}).  A set of ODE's whose solutions satisfy
these criteria is easily found to be~\cite{dixon}
\begin{eqnarray}\label{eq:circles}
\dot{x} & = & y\nonumber\\
\dot{y} & = & \frac{y^2}{2 x} - \frac{1}{2}x.
\end{eqnarray}

We will not examine Eqs.\ (\ref{eq:circles}) in detail except to note that
while the origin is a common point of all dynamical trajectories,
it is not a fixed point.  A particular circle of radius $r$ (for positive
$x$) is given by
\begin{equation}
(x - r)^2 + y^2 = r^2,
\end{equation}
or,
\begin{equation}
y^2 = 2 x r - x^2.
\end{equation}
The equation for $\dot{y}$ may then be re-expressed as
\begin{equation}
\dot{y} = \frac{2 x r - x^2}{2 x} - \frac{1}{2}x
 = r - x.
\end{equation}
In the limit that $x,y \rightarrow 0$, we find simply that $\dot{y} = r$.
Excepting the case of the trivial circle ($r=0$), we see that the origin
is not a fixed point and any dynamical trajectory will reach the origin
in finite time.\\

Let us now add some very small random fluctuations to our system.
Equations\ (\ref{eq:circles}) then become
\begin{eqnarray}\label{eq:circnoise}
\dot{x} & = & y + \epsilon f(t)\nonumber\\
\dot{y} & = & \frac{y^2}{2 x} - \frac{1}{2}x + \epsilon g(t),
\end{eqnarray}
where $\epsilon \ll 1$ and $f(t)$ and $g(t)$ are identically distributed
random functions (we have named them differently to denote that they
take on different functional values for the same $t$).  Usually, if
$\epsilon$ is small enough, we ignore such fluctuations, as they
have little effect on the dynamics.  Indeed, as long as the dynamical
trajectory is away from the origin, the average solution of
Eqs.\ (\ref{eq:circnoise}) will correspond to a solution of the
unperturbed Eqs.\ (\ref{eq:circles}).  Near the origin, however,
such fluctuations may have a dramatic effect, since any small change
in $x$ and $y$ could land on a circle with {\em any} radius ranging
from $0$ to $\infty$.\\

The dynamics of Eqs.\ (\ref{eq:circnoise}) are then expected to be similar
to those of the RKHO.  The solution
(on average) moves along a circle of particular $r$ until it
reaches some neighborhood about the origin where the fluctuations
become prevalent.  The trajectory will leave this neighborhood on a circle
of some different $r$ which was randomly selected by the action of
the fluctuations.  Like the RKHO, we see that circles before and
after the origin are uncorrelated, and that the dynamics is correspondingly
unpredictable.  Unlike the RKHO, the size of the fluctuations may be
arbitrarily small, yet the resultant behavior will be equally
non-deterministic (in essence, the random kick has been replaced
by the infinite divergence at the origin).  We term this dynamical behavior
{\em non-deterministic chaos}.  The key feature of the underlying
dynamical system is a point such as the origin for
Eqs.\ (\ref{eq:circles}), which is reachable
in finite time, but is a common point among many dynamical trajectories.
We denote such a point as a {\em sensitive decision point} (SDP).\\

It should be noted here that the type of behavior described above is
not unique to the particular conditions we have put forth.  The case
of {\em noise induced instability}, as described by Chen~\cite{chen},
also produces large-scale stochastic behavior in the presence of small
random fluctuations.  The underlying causes, however, are somewhat
different.  Noise induced instability occurs when the random perturbations
drive the dynamical trajectory over the potential barrier created by
a driving force~\cite{chen}.  Non-deterministic chaos is essentially
the result of a singularity in the equations of motion (i.e., infinite
divergence at a point), which allows dynamical orbits comprising a finite
region of the phase plane to intersect at a single point.\\

\paragraph*{Statistical Properties}

In the case of the RKHO, the statistics of the orbit were obviously
dependent on the statistics of the driving function.  This is not
so obvious in the case at hand, but we should be able to make some
statistical predictions about the dynamics.  More precisely, we would
like to know the probability that a circle of given $r$ is chosen when
the orbit leaves some neighborhood about the origin.  Let us define
this neighborhood as a disk of radius $\delta$, and note that an
orbit leaving this neighborhood does so with angle $\theta$, which
we take as measured from the $y$-axis.  Given that we expect the
fluctuations to be isotropic, the probability density of picking
a particular $\theta$ is constant, i.e.,
\begin{equation}
p(\theta)d\theta \propto d\theta.
\end{equation}
Next we note that everywhere
except at the origin the Existence and Uniquenss Theorem applies to
solutions of Eqs.\ (\ref{eq:circles}), thus each circle of radius
$r$ is associated with a unique $\theta$, and we may write $\theta$
as a function of $r$.  Substituting into $p(\theta)$, we find
\begin{equation}
p(r)dr \propto \frac{\partial \theta(r)}{\partial r}dr.
\end{equation}
The probability of getting a circle between $r$ and $r + \Delta r$
is simply
\begin{equation}
P(r,\Delta r) \propto
\int_{r}^{r+\Delta r}\frac{\partial \theta(r)}{\partial r}dr
 = \theta(r+\Delta r) - \theta(r).
\end{equation}
For the case at hand, we find
\begin{equation}
P(r,\Delta r) \propto \arccos{\frac{\delta}{2(r+\Delta r)}}
			- \arccos{\frac{\delta}{2r}}.
\end{equation}

\paragraph*{Discussion}

To date, there is one candidate for a physical non-deterministic chaotic
system.  This is a model of neutron star dynamics~\cite{cummings}
which exhibits a sensitive decision point and is expected to behave
as described above~\cite{dixon,dixon2}.  Whether other physical
examples of non-deterministic chaos exist is not clear at this time.
However, it is known that apparently complex systems may often
successfully be modeled by a low-dimensional deterministic chaotic
system~\cite{rasband}.
We cautiously advance the possibility of an analogous situation for
non-deterministic chaos.  This is attractive, since the analysis
and modeling of a non-deterministically chaotic signal is reduced to
finding the SDP's, doing some curve fitting of the signal between
SDP's, and performing some straightforward statistical analysis as
described above.  We are continuing to study such possibilities.\\

\begin{figure}
\caption{Some examples of circular orbits of different radii, all sharing
a common point at the origin.}
\label{fig:circles}
\end{figure}

\end{document}